\documentstyle[12pt]{article}

\newcommand{\be}{\begin{equation}}
\newcommand{\ee}{\end{equation}}
\newcommand{\bea}{\begin{eqnarray*}}
\newcommand{\eea}{\end{eqnarray*}}
\newcommand{\ba}{\begin{eqnarray}}
\newcommand{\ea}{\end{eqnarray}}

\begin{document}

\begin{flushright} UCL-IPT-95-10 \end{flushright}

\thispagestyle{empty}

\vspace*{20mm}

\begin{center}
\Large{\bf The asymptotic behaviour of the $\pi^0$ $\gamma^\ast$
$\gamma^\ast$ vertex.}
\end{center}

\vspace{20mm}

\begin{center}
J.-M. G\'erard and T. Lahna
\end{center}

\vspace{20mm}

\begin{center}
Institut de Physique Th\'eorique\\
Universit\'e catholique de Louvain\\
B-1348 Louvain-la-Neuve
\end{center}

\vspace*{5mm}

\begin{center}
{\large To the memory of Roger Decker}
\end{center}

\vspace{20mm}

\begin{abstract}
The Bjorken-Johnson-Low theorem applied to the $\gamma^\ast \to \gamma^\ast
\pi^0$
process provides us with a rather remarkable asymptotic behaviour for the
$\pi^0
\gamma^\ast \gamma^\ast$ vertex. We compare our result with previous QCD-
inspired
estimates and argue that the predicted behaviour is quite consistent with
the present
data on hadronic $J/\Psi$ decays and ensures the matching of long- and
short-distance
radiative corrections to
$\pi^+
\to e^+ \nu_e$.
\end{abstract}

\newpage

Anomalies play a crucial role in elementary particle physics. Their
cancellation for
local currents \cite{1} to ensure the renormalizability of the (extended)
Standard
Model based on the $SU(N)_C \times SU(2)_L \times U(1)$ gauge symmetries
requires a
relation between the electric charge and the (odd) number $N$ of colours for
the
up and down quarks:
\ba
Q_u &=& \frac{1+N}{2N} \nonumber\\
& & \label{1}\\
Q_d &=& \frac{1-N}{2N}.\nonumber
\ea
On the other hand, the non-vanishing triangle anomaly \cite{2} associated
with a global
axial current allows the $\pi^0$ to decay into two real photons at a rate
in excellent
agreement with experiment if its fermionic components precisely satisfy the
relations
given in Eq.(\ref{1}), for any $N$. In particular, for $N=1$ we recover the
old result
\cite{3} based on a nucleon bound state picture for the pion and for $N=3$,
the standard one \cite{4} based on the quark model for $\pi^0$.

It is therefore quite interesting to study extrapolations of the anomalous
$\pi^0
\gamma \gamma$ vertex to virtual photons. From a phenomenological point of
view, the
$W^+ \to \pi^+ \gamma$ and $Z^0 \to \pi^0 \gamma$ processes \cite{5} are
related to
this vertex with one off-mass-shell photon, while the two off-mass-shell
photon vertex
contributes to the rare $\pi^0 \to e^+ e^-$ decay amplitude \cite{6}.

Different asymptotic behaviours for the $\pi^0 \gamma^\ast (q^2_1)
\gamma^\ast (q^2_2)$
vertex when $(q^2_1 \to \infty , q^2_2 = 0)$ \cite{7} or $(q^2_1 = q^2_2
\to \infty)$
\cite{8} have been derived in the literature. On the sole basis of the
Bjorken-Johnson-Low theorem \cite{9}, we will argue that the asymptotic
behaviour for
$(q^2_1
\to
\infty, q^2_2$ arbitrary) is in fact universal. This result is compatible
with the
present data on hadronic decays of the $J/\Psi$. Moreover it implies a
smooth matching
between the short-distance and long-distance electromagnetic corrections to
the $\pi^+
\to e^+ \nu_e$ weak process.

\vspace{8mm}
\par\noindent
{\bf 1.} {\bf The $\gamma^\ast \to \gamma^\ast \pi^0$ transition and the
Bjorken-Johnson-Low theorem.}

The Bjorken-Johnson-Low (BJL) theorem \cite{9} can be used to derive the
matrix element for the $\gamma^\ast \to \gamma^\ast \pi$ process when one
photon is very energetic.

If $(q_1,\epsilon_1)$ and $(q_2,\epsilon_2)$ are the momentum and
polarization of
the incoming and outgoing photons respectively, the amplitude for
$\gamma^\ast \to
\gamma^\ast
\pi$ is given by
$$
M = M_{\mu\nu} \;\; \epsilon^\mu_1 \;\; \epsilon^\nu_2 $$
where
\ba
M_{\mu\nu} &=&
i e^2 \int d^4 x \;\; e^{iq_2 x} \langle \pi^0 | T \{ J_\nu (x) J_\mu (0) \} |
0
\rangle
\nonumber \\
&=&
ie^2 \int d^4 x \;\; e^{iq_1 x} \langle \pi^0 | T \{ J_\mu (-x) J_\nu (0) \} |
0
\rangle.
\label{2}
\ea
In the limit $q^0_1 \to \infty$, the well-known BJL theorem \cite{9}
applied to electromagnetic currents implies
\be
M_{\mu\nu} (q^0_1 \to \infty) = - \frac{e^2}{q^0_1} \int d \vec x \;\;
e^{-i \vec q_1
\vec x}
\langle \pi^0 | [ J_\mu (0,-\vec x\,) , J_\nu (0,0) ] | 0 \rangle . \label{3}
\ee

If we restrict ourselves to the two-flavour quark model, the
electromagnetic current
$J_\mu$ reads
\be
J_\mu = \bar \Psi \gamma_\mu Q \Psi
\label{4}
\ee
with
$$
\Psi = \left(\begin{array}{c} u \\ d \end{array} \right) \;\;\;\;\; ,
\;\;\;\;\;
Q = \left(\begin{array}{ccccc} Q_u & 0 \\ 0 & Q_d \end{array} \right). $$
Using the canonical equal time anticommutation relations for the quark
fields \be
\{ \bar \Psi^\alpha_i (0,- \vec x\,) , \Psi^\beta_j (0,0) \} = \gamma_0
\;\; \delta_{ij}\;\;
\delta^{\alpha\beta} \;\; \delta (-\vec x\,) \label{5}
\ee
with $i,j = 1...N$ and $\alpha,\beta=u,d$, the colour and flavour indices
respectively, we obtain the equal-time commutator \be
[J_\mu (0,-\vec x\,),J_\nu (0,0)] = 2 i \;\; \epsilon_{\mu\nu0\sigma} \;\;
\delta
(-\vec x\, ) \;\; J^\sigma_5 (0)
\label{6}
\ee
in terms of the axial-vector current
\be
J^\sigma_5 = \bar \Psi \gamma^\sigma \gamma_5 Q^2 \Psi. \label{7}
\ee
{}From the pion-to-vacuum matrix element \be
\langle 0 | \bar u \gamma^\sigma \gamma_5 d | \pi^- (p) \rangle \equiv i
\;\; f_\pi \;\; p^\sigma
\label{8}
\ee
with $f_\pi \sim$ 132 MeV, the $\pi \to \mu\nu$ decay constant, we infer
that \be
M_{\mu\nu} (q^0_1 \to \infty) = e^2 (Q^2_u - Q^2_d) \frac{\sqrt 2 f_\pi}{q^0_1}
\epsilon_{\mu\nu 0 \sigma} (q_2 - q_1)^\sigma. \label{9}
\ee
If we multiply and divide the right-hand side of Eq.(\ref{9}) by $q^0_1$
and assume
$|{\vec q}_1
\, | << q^0_1$, we obtain
\be
M_{\mu\nu} (q^2_1 \to \infty) = e^2 (Q^2_u - Q^2_d) \sqrt 2
\frac{f_\pi}{q^2_1} \epsilon_{\mu\nu\rho\sigma} q^\rho_1 q^\sigma_2.
\label{10}
\ee
The standard $\pi^0 \gamma (q^2_1) \gamma (q^2_2)$ form factor
$F(q^2_1,q^2_2)$ is
defined as
\be
M_{\mu\nu} (q^2_1,q^2_2) \equiv e^2 F(q^2_1,q^2_2)
\epsilon_{\mu\nu\rho\sigma} q^\rho_1 q^\sigma_2.
\label{11}
\ee
With this normalization, the $\pi^0 \to \gamma \gamma$ triangle anomaly
\cite{2}
implies
\be
F(q^2_1 = 0 , q^2_2 = 0) = - \frac{\sqrt 2}{4\pi^2} \frac{1}{f_\pi}. \label{12}
\ee
On the other hand, from Eq.(\ref{1}) and Eq.(\ref{10}) based on the BJL
theorem we
obtain the following asymptotic behaviour for the $\pi^0 \gamma^\ast
\gamma^\ast$
vertex:
\be
F(q^2_1 \to \infty, q^2_2 \; \mbox{finite}) = + \frac{\sqrt 2}{N}
\frac{f_\pi}{q^2_1} .
\label{13}
\ee
This result has been obtained for time-like photon momenta. One can easily
show that it
remains valid in the space-like region.

Since
\be
f_\pi \sim \sqrt N
\label{14}
\ee
the dependence on the number of colours
is the same in Eqs (\ref{12}) and (\ref{13}), as it should be in a quark
model for the
pion.

The remarkable feature of the main result displayed in Eq.(\ref{13}) is its
validity
for any finite value of $q^2_2$. Such a feature is not expected from
QCD-inspired
derivations.

\vspace{8mm}
\par\noindent
{\bf 2.} {\bf A comparison with previous QCD estimates of $F(q^2_1,q^2_2)$.}

The Operator Product Expansion approach (OPE) has been used \cite{8} for
the symmetric
case of two highly energetic photons with equal square momenta $(q^2_1 = q^2_2
=
q^2)$. The resulting form-factor
\be
F^{OPE} (q^2 \to \infty, q^2 \to \infty) = \frac{\sqrt 2}{N} \;
\frac{f_\pi}{q^2}
\label{15}
\ee
has very small QCD corrections and nicely extends the asymptotic behaviour
given in
Eq.(\ref{13}) to arbitrary values of $q^2_2$. This OPE approach applied to
the case of
one real photon
$(q^2_2 = 0)$ implies \cite{10}
\be
F^{OPE} (q^2_1 \to \infty , q^2_2 = 0) = \frac{2\sqrt 2}{N} \;
\frac{f_\pi}{q^2_1}
\;\;\;\; + \; \mbox{large corrections}
\label{16}
\ee
with potentially large QCD corrections. For the same reason, the
QCD-inspired calculation of Lepage and Brodsky (LB)
\cite{7}
\be
F^{LB} (q^2_1 \to \infty , q^2_2 = 0) = \frac{3\sqrt 2}{N}
\frac{f_\pi}{q^2_1} \label{17}
\ee
cannot be trusted \cite{10}.

Predictions \cite{5} based on
the LB form-factor given in Eq.(\ref{17}) for processes like $Z^0 \to \pi^0
\gamma$ or
$W^+
\to \pi^+ \gamma$ are therefore highly questionable. It is however unlikely
to get any
information on the form factor $F(q^2_1,0)$ from these very rare processes
in a near
future. But present data on hadronic $J/\Psi$ decays provide already a way
to discriminate in favour of Eq.(\ref{13}) with relatively small
corrections.

\vspace{8mm}
\par\noindent
{\bf 3.} {\bf The $J/\Psi \to \omega \pi^0$ decay and $F(q^2_1 \approx 9.6$
Gev$^2$,
$q^2_2 = 0$).}

Among the measured $J/\Psi$ hadronic two-body decays into one vector and one
pseudoscalar, the $J/\Psi \to \omega \pi^0$ is the only one to be clearly
dominated by the single-photon exchange contribution $(Q_u \neq Q_d)$.
Indeed, the
contributions induced by the isospin-violating $\eta-\pi^0$ mixing $(m_u
\neq m_d)$ in
the $J/\Psi
\to \omega \eta$ decay and $\rho-\omega$ mixing ($m_u \neq m_d)$ in the
$J/\Psi \to
\rho \pi^0$ decay are negligible. Notice that the other isospin-violating
$J/\Psi$ decays into $\rho \eta$ and $\rho \eta'$ involve the strong
anomaly. Moreover, they are contaminated by $\omega-\rho$ mixing $(m_u \neq
m_d)$ in the
$J/\Psi \to \omega \eta, \omega \eta'$ decays since the $\omega$ width is
small.

{}From the measured $J/\Psi \to \omega \pi^0$ decay width \cite{11} \ba
\Gamma (J/\Psi \to \omega \pi^0) &=& (3.7 \pm 0.5) 10^{-8} \; \mbox{GeV}
\nonumber \\
&\simeq& \frac{1}{96 \pi} m^3_\Psi g^2_{\Psi \omega \pi^0}
[1-\Bigg(\frac{m_\omega}{m_\Psi}\Bigg)^2]^3 \label{18}
\ea
we estimate the $\Psi \omega \pi^0$ vertex $g_{\Psi\omega\pi^0}$: \be
g_{\Psi\omega\pi^0} \simeq (6.8 \pm 0.5) 10^{-4} \; \mbox{GeV}^{-1}. \label{19}
\ee
The dominance of the single-photon exchange for this specific process
implies then the simple relation
\be
e^2 F (m^2_\Psi,0) = \frac{f_\Psi}{f_\omega} g_{\Psi \omega \pi^0} \label{20}
\ee
if the usual vector-dominance model is assumed for the $\omega-\gamma$
transition.
The decay constant $f_V$ of the vector-meson $V$ is defined such that \be
\langle 0 | J_\mu | V \rangle = \epsilon_\mu \frac{m^2_V}{f_V}. \label{21}
\ee
With this normalization, we obtain $f_\omega \simeq 17.1$ and $f_\Psi
\simeq 11.5$
from the central values of the measured $\omega \to e^+e^-$ and $\Psi \to
e^+e^-$ decay
widths
\cite{11} given by the general expression \be
\Gamma (V \to e^+e^-) = \frac{\alpha^2}{3} \frac{4\pi}{f^2_V} m_V. \label{22}
\ee
Keeping in mind the various approximations used, we get a reasonable
estimate of the
$\pi^0 \gamma^\ast (q^2_1 = m^2_\Psi) \gamma (q^2_2 = 0)$ from factor \be
e^2 F(m^2_\Psi,0) = (4.6 \pm 0.4) \;\;\; 10^{-4} \; \mbox{GeV}^{-1}. \label{23}
\ee

On the other hand, assuming $q^2_1 = m^2_\Psi$ in the asymptotic expression
given
in Eq.(\ref{13}) would imply
\be
e^2 F(m^2_\Psi,0) \to e^2 \frac{\sqrt{2}}{N} \frac{f_\pi}{m^2_\Psi} = 6
\;\;\; 10^{-4} \; \mbox{GeV}^{-1}
\label{24}
\ee
for the standard value $N=3$. This prediction is in good agreement with the
one extracted from experiment in Eq.(\ref{23}).

We conclude that the observed $J/\Psi \to \omega \pi^0$ hadronic decay rate
alone
suggests that the $J/\Psi$ mass scale is already in the asymptotic domain
where the
BJL theorem applies. It also indicates the failure of the leading order QCD
estimates
given in Eqs (\ref{16}) and (\ref{17}) which predict $e^2F(m^2_\Psi,0)
\simeq 12 \;\;
10^{-4}$ GeV$^{-1}$ and $18
\;\; 10^{-4}$ GeV$^{-1}$ respectively. A crucial test to settle the
question would be
the measurement of the
$\Upsilon \to \omega \pi^0$ hadronic decay rate since the BJL theorem
should be valid
at this scale. Assuming Eq.(\ref{13}), we estimate the $\Upsilon \omega
\pi^0$ vertex to be
\be
g_{\Upsilon \omega \pi^0} \simeq \frac{f_\omega}{f_\Upsilon} e^2
\frac{\sqrt 2}{N}
\frac{f_\pi}{m^2_\Upsilon}.
\label{25}
\ee
{}From the measured $\Upsilon \to e^+e^-$ decay width \cite{11}, we have
$f_\Upsilon
\approx$ 40 (see Eq.(\ref{22})). For $N=3$, the predicted $\Upsilon \to
\omega \pi^0$
branching ratio is about
\be
Br (\Upsilon \to \omega \pi^0) \approx 3.8 \;\;\;\; 10^{-5}. \label{26}
\ee
The LB form factor in Eq.(\ref{17}) would imply a branching ratio larger by
a factor of
nine ! Unfortunately, data on this process are not available yet \cite{11}.

We have seen that a direct test of the BJL theorem applied to the
$\gamma^\ast \to
\gamma^\ast \pi^0$ transition is still lacking. Now we present a
theoretical argument
in favour of the general asymptotic behaviour in Eq.(\ref{13}).

\vspace{8mm}
\par\noindent
{\bf 4.} {\bf Matching in the radiative corrections to $\pi^+ \to e^+ \nu_e$.}

The one-loop electromagnetic corrections to the zeroth-order weak $\pi^+ \to
e^+
\nu_e$ decay amplitude ${\cal M}_0$ are well-known \cite{12}. As long as
the $q^2_2$ square momentum of the charged $W$ gauge boson is negligible
with respect to
$M^2_W$, these corrections denoted $\delta M_0$ can be classified according
to the
$q^2_1$ square momentum carried by the virtual photon.

For $q^2_1 > \Lambda^2$, one uses a quark model for the charged pion and
considers
the short-distance (SD) corrections to the generic $u \bar d \to e^+ \nu_e$
process. Using Eq.(\ref{1}), the leading dependence on the infrared
euclidean cut-off
$\Lambda$ reads
\cite{13}
\be
\delta {\cal M}^{SD}_0 = [ \frac{3\alpha}{8\pi} (1+\frac{1}{N}) \ln
\frac{M^2_W}{\Lambda^2}] {\cal M}_0.
\label{27}
\ee

For $q^2_1 < \Lambda^2$, one treats the pion as an elementary pseudoscalar
and considers the long-distance (LD) corrections to $\pi^+ \to e^+ \nu_e$.
The leading
dependence on the ultraviolet euclidean cut-off $\Lambda$ reads \cite{14} \be
\delta {\cal M}^{LD}_0 = [ \frac{3\alpha}{8\pi} \ln
\frac{\Lambda^2}{m^2_\pi}]{\cal M}_0.
\label{28}
\ee

The cut-off $\Lambda$ introduced by hand to separate short-distance from
long-distance radiative corrections around the GeV scale is obviously
unphysical.
Consequently, the total $\delta {\cal M}_0$ corrections to the $\pi^+ \to
e^+ \nu_e$
weak amplitude should be $\Lambda$-independent. This smooth matching
between the two
complementary pictures for the pion has been successfully implemented in
the case of
the $\pi^+ - \pi^0$ mass difference \cite{15}.

Here, we might prematurely conclude from Eqs (\ref{27}) and (\ref{28}) that
this
matching only occurs in the large-$N$ limit, when $Q_u = - Q_d = 1/2$. In this
limit indeed, the $1/N$ term in Eq.(\ref{27}) arising from the {\em vector}
component of the weak hadronic current coupled to $W^-$ disappears and the
matching
between the axial component contributions occurs.

However, we have seen that, for arbitrary $N$, the structure of the pion
coupled to
two vector currents implies a $ 1/N$-suppressed form-factor whose
asymptotic behaviour is given in Eq.(\ref{13}). This also applies for the
isospin related $\pi^+
\gamma^\ast (q^2_1) W^- (q^2_2)$ vertex appearing in the long-distance
radiative
corrections. Integrating over large photon momenta, we obtain: \be
\delta {\cal M}^{structure}_0 = [\frac{3\alpha}{8\pi} \int^{\Lambda^2}
dq^2_1 \;\;\;\; F(q^2_1,q^2_2 < M^2_W)] \frac{{\cal M}_0}{\sqrt 2 f_\pi}.
\label{29}
\ee
We emphasize that the general asymptotic behaviour advocated in
Eq.(\ref{13}) is valid for finite $q^2_2$ and ensures therefore a complete
cancellation of the
leading cut-off dependence in
$\delta {\cal M}_0 =
\delta {\cal M}^{SD}_0 +
\delta {\cal M}^{LD}_0 + \delta {\cal M}^{structure}_0$, for any value of $N$.

As expected from PCAC applied on $\pi^0$, we reach the same conclusion in
the case of
the electromagnetic corrections to the
$\pi^+ \to \pi^0 e^+ \nu_e$ process. For this pion $\beta$-decay, CVC
automatically
ensures the matching between the short-distance and long-distance
contributions induced by the vector component of the weak hadronic current.
The short-distance
logarithmic dependence on the cut-off $\Lambda$, arising now from the {\em
axial}
component of the weak hadronic current, is then cancelled by the
$\pi^+\pi^0 \gamma^\ast W^-$ vertex. The asymptotic behaviour of this
vertex is again simply
deduced from the BJL theorem and was already known about thirty years ago
\cite{16} !

In conclusion, we have shown how the Bjorken-Johnson-Low theorem provides
us with a
unique asymptotic behaviour for the $\pi^0 \gamma^\ast (q^2_1)\gamma^\ast
(q^2_2)$
vertex when $|q^2_1|$ tends to infinity. This surprising result was not
expected from
previous QCD-inspired approaches. We then argued that present data on
$J/\Psi$ hadronic decays support this result and emphasized the relevance
of measuring the
$\Upsilon
\to \omega \pi^0$ branching ratio. Finally, we explained how the matching
between
short-distance and long-distance electromagnetic corrections to $\pi^+ \to
e^+ \nu_e$
is naturally implemented in this framework.

\vspace{20mm}

\par\noindent
{\Large{\bf Acknowledgements}}

We would like to thank Roger Decker, Jean Pestieau and Jacques Weyers for
very useful
discussions and comments.

\end{document}